\documentclass[]{iopart}

\usepackage{amssymb} 

\newcommand{\Exp}[1]{\,\mathrm{e}^{\mbox{\footnotesize$#1$}}}
\newcommand{\I}{\mathrm{i}}

\newenvironment{eqnArray}{\begin{equation}\begin{array}[b]{rcl}}
{\end{array}\end{equation}}
\newcommand{\ds}{\displaystyle}
\newcommand{\Par}[1]{\mathrm{Par(#1)}}

\begin{document}

\title[Symmetric coupling of four spin-1/2 systems]%
{Symmetric coupling of four spin-1/2 systems}
\author{Jun Suzuki$^1$ and Berthold-Georg Englert$^{2,3}$}

\address{$^1$ %
Graduate School of Information Systems, %
The University of Electro-Communications, %
1-5-1 Chofugaoka, Chofu, Tokyo 182-8585, Japan}
\address{$^2$ %
Centre for Quantum Technologies, %
National University of Singapore, Singapore 117543, Singapore}
\address{$^3$ %
Department of Physics, %
National University of Singapore, Singapore 117542, Singapore}

\ead{junsuzuki@is.uec.ac.jp, cqtebg@nus.edu.sg}

\begin{abstract}
We address the non-binary coupling of identical angular momenta 
based upon the representation theory for the symmetric group. 
A correspondence is pointed out between the complete set of commuting operators 
and the reference-frame-free subsystems.  
We provide a detailed analysis of the coupling of three and four spin-1/2 
systems and discuss a symmetric coupling of four spin-1/2 systems. 
\end{abstract}

\pacs{03.65.-w,03.65.Fd}  
\ams{20B35}               

\renewcommand{\submitto}[1]{\vspace{28pt plus 10pt minus 18pt}
     \noindent{\small\textrm{Posted on the arXiv on #1}}}
\submitto{12 April 2012}

\section{Introduction}\label{sec:Intro}
The coupling of several angular momenta is one of the important technical
problems in quantum mechanics. 
Every standard textbook of quantum mechanics discusses at least the coupling
of two angular momenta. 
This problem is of great importance not only for practical purposes when
dealing with several particles such as in atomic physics, nuclear physics,
and so on, but is also of quite some mathematical interest.
To be specific, the problem can be rephrased as a well-known problem in group 
representation theory, namely to obtain a direct sum of irreducible
representations of the rotation group. 
There is a large body of literature in which this problem is studied 
both analytically and numerically; 
see, for example, \cite{AMinQP1, AMinQP2}. 

The complete analytical formula is known for the coupling of any two 
angular momenta in terms of the Clebsh-Gordan (CG) coefficients. 
These coefficients immediately allow us to express the bases in irreducible
representations as super\-positions of the bases in direct product
representations.  
It is a well established strategy to add more than two angular momenta 
by successive applications of the addition of two with the aid of CG
coefficients.  
The addition of many angular momenta can be carried out by using the $nj$
symbols, the Racah coefficients, and so on, which are generally 
studied within the recoupling theory. 
Indeed, the study of many angular momenta turns out to be a beautiful
mathematical physics problem in its own right. 

Recently, the study of irreducible representations of many identical spin-$j$
systems, or many $d$-dimensional quantum systems (qudits), has been revived in
the context of quantum information and quantum computing. 
When coupling many spin-$j$ systems, the structure of the Hilbert space can be
described by representation theory of the symmetric group. 
This fact is known as the Weyl-Schur duality which can be utilized to solve
many problems~\cite{GW}. 
Interesting examples of this kind include 
an estimation for a spectrum of an unknown quantum state~\cite{KW}, 
quantum communication without sharing a reference frame~\cite{BRS}, 
universal coding for a classical-quantum channel~\cite{hayashi}, 
and others~\cite{more}. 
Another interesting feature among these studies is the proposal for an
efficient quantum circuit to obtain the irreducible representation of the
$N$-fold tensor product of a $d$-dimensional Hilbert space, which requires
only a total number of gates of order $N\log(d,\log N, \log 1/\epsilon)$ 
up to accuracy $\epsilon$~\cite{BCH1,BCH2,Harrow}. 

Let us look at the coupling of several angular momenta using the CG
coefficients.  
The first step is to add two angular momenta which are conveniently chosen from 
all angular momenta. 
The next step is then to add each of the obtained angular momenta 
and another one chosen from the yet-uncoupled angular momenta. 
By repeating this binary coupling many times, one can arrive at the desired
result.  
It is a rather straightforward task to perform each step, but the final result
cannot be obtained in a simple manner. 
The major obstacle is that the computational complexity 
of such a coupling of several angular momenta grows rather rapidly with
the total number of angular momenta and the dimension of each
angular momentum.  

As the simplest case, we consider the addition of $N$ spin-1/2 systems which
can be decomposed into a direct sum of irreducible representations labeled
with angular momentum $j$ as  
\begin{equation} \label{decom}
\mathcal{D}_{1/2}^{\otimes N}=\bigoplus_{j \in J}c_j\mathcal{D}_j,
\end{equation}
where the index set $J=\{N/2,N/2-1,\dots\}$ has $(N+1)/2$ or $N/2+1$ elements
if $N$ is odd or even, respectively. 
The multiplicity of each irreducible space is 
\begin{equation} \label{mult}
c_j=\frac{N!(2j+1)}{(N/2+j+1)!(N/2-j)!}.
\end{equation} 
When another spin-1/2 system is to be added to the obtained result, we need to 
couple this new spin-1/2 state to $\sum_j c_j\sim(2N)!/(N!)^2\sim 4^N$
different angular momenta, which number grows exponentially for large $N$.

Another disadvantage of the binary coupling is that the various constituent
angular momenta are not treated on equal footing. 
In other words, the resulting angular momentum states depend on the way one
chooses the paring in the intermediate steps.  
The binary coupling might not be a wise choice when dealing with many identical 
angular momenta. 
To overcome this problem, a novel coupling scheme was proposed more than four
decades ago independently by Chakrabarti~\cite{sym1} and L\'evy-Leblond  
and L\'evy-Nahas~\cite{sym2}. 
They studied the non-binary couplings of three angular momenta 
without employing the binary coupling. 
Their coupling scheme is generally referred to as the symmetric coupling or
the democratic coupling, which reflect the fact that  
their choice of the complete set of commuting operators (CSCO) contains 
all three angular momenta with equal weights. 
To our knowledge, there has been no generalization of their non-binary
coupling to the case of more than three angular momenta. 

It is our main motivation here to clarify the meaning of the symmetric coupling 
for four identical spin systems and then to provide a possible solution for
the coupling of four spin-1/2 systems. 
While the main result was already reported in \cite{JGB}, 
we have not clarified the meaning of the symmetric coupling as yet. 
We hope that our construction paves the way toward establishing a non-binary
coupling of many angular momenta.  

This paper is organized as follows. 
In section~\ref{sec:2} we give the mathematical background 
of the Weyl-Schur duality, a possible definition of the symmetric coupling of 
many identical angular momentum systems, and discuss the relation to the 
reference-frame-free subsystems. 
We then study the symmetric coupling of many spin-1/2 systems in the second
largest angular momentum subspace in section~\ref{sec:3}.  
The detailed analysis in the case of four spin-1/2 systems is shown in
section~\ref{sec:4}.  
We close with a summary and discussion in section~\ref{sec:SummDisc}

\section{Symmetric coupling of $N$ spin-1/2 systems}\label{sec:2}
We briefly summarize some of relevant mathematical facts. 
Readers are referred to \cite{GW} for more concise discussions. 
In the rest of paper, we mainly consider $N$ spin-1/2 systems 
(2-dimensional systems) unless stated explicitly.

\subsection{Weyl-Schur duality}
The $N$-fold tensor product of the $2$-dimensional Hilbert space can be
decomposed into the following direct sum structure:  
\begin{equation} \label{WSdual}
(\mathbb{C}^2)^{\otimes N}=\bigoplus _{\nu\in\Par{N,2}}
\mathcal{S}_{\nu}\otimes \mathcal{R}_{\nu} ,
\end{equation}
where $\Par{N,2}$ stands for the partition of $N$ into two
non-negative and non-increasing integers,  i.e., 
${\Par{N,2}=\{(\nu_1,\nu_2)\in\mathbb{Z}^2|%
\nu_1+\nu_2=N,\nu_1\geq\nu_2\geq0\}}$. 
In the decomposition (\ref{WSdual}), known as the Wedderburn decomposition, 
the subspaces $\mathcal{R}_{\nu}$ are the representation spaces for the
general matrix group over the complex field with the dimension 
\begin{equation}
r(\nu)=\nu_1-\nu_2+1,
\end{equation}
and $\mathcal{S}_{\nu}$ are the representation spaces for the symmetric group
$S_N$ with the dimension  
\begin{equation}
s(\nu)=\frac{N!\,(\nu_1-\nu_2+1)}{(\nu_1+1)!\,\nu_2!}.
\end{equation} 
This dimension $s(\nu)$ is same as the multiplicity $c_j$ in (\ref{mult}) for
$j=(\nu_1-\nu_2)/2$. 
In other words, we can also label the subspaces with a single quantum number
$j$ in accordance with 
\begin{equation}
\nu_{1}=\frac{N}{2}+j,\quad
\nu_{2}=\frac{N}{2}-j, 
\end{equation}
with ${j\in J}$. 
Note that $r(\nu)=2j+1$ is the dimension of the subspace $\mathcal{D}_j$ with
angular momentum $j$.  
As an example, consider the coupling of three spin-1/2 systems, in which case
the partition is $\Par{3,2}=\{ (3,0), (2,1) \}$. 
The dimensions of the corresponding subspaces are 
$r(3,0)=4$, $s(3,0)=1$, $r(2,1)=2$, and $s(2,1)=2$. 

Because of the decomposition (\ref{WSdual}), the $N$-fold tensor product of
the two-dimensional non-singular matrices $A\in\mathrm{GL}(2,\mathbb{C})$, 
i.e., $A^{\otimes N}$, acts irreducibly on the subspaces $\mathcal{R}_{\nu}$, and 
the unitary representations of the permutation operators $P_{i_1 i_2 \dots i_N}$ 
act irreducibly on the subspaces $\mathcal{S}_{\nu}$. 
Throughout the paper, we denote the permutation from $1,2,\dots, N$ to $i_1
i_2 \dots i_N$ by $P_{i_1 i_2 \dots i_N}$, that is, 
\begin{equation}
P_{i_1 i_2 \dots i_N}\equiv
\left(\begin{array}{cccc}
1 & 2 & \dots & N \\i_{1} & i_{2} & \dots & i_{N}
\end{array}\right). 
\end{equation}
Therefore, $A^{\otimes N}$ and $P_{i_1 i_2 \dots i_N}$ are decomposed as follows:
\begin{eqnArray} \label{WSdual2}
A^{\otimes N}&=&
\ds\bigoplus _{\nu\in\Par{N,2}} I_{s(\nu)}\otimes R_{\nu},\nonumber\\[4ex] 
\label{WSdual3}
P_{i_1 i_2 \dots i_N}&=&
\ds\bigoplus _{\nu\in\Par{N,2}} S_{\nu}\otimes I_{r(\nu)}.
\end{eqnArray}
These decompositions are the essence of the Weyl-Schur duality which states
that operators commuting with all elements of $A^{\otimes N}$ are expressed as
linear combinations of the unitary permutation operators 
$P_{i_1 i_2 \dots i_N}$ with complex coefficients.  
Moreover, its inverse also holds, that is, if operators commute with all
elements of $P_{i_1 i_2 \dots i_N}$, they are a linear combination of
$A^{\otimes N}$ with complex coefficients.  
The Weyl-Schur duality holds for the general case of $N$-fold tensor products
of $d$-dimensional systems~\cite{GW}.

\subsection{Complete set of commuting operators and missing label operators}
As a mathematical problem, to calculate the completely reducible
representation is equivalent to finding the CSCO whose joint eigenstates
define the representation uniquely up to arbitrary phase factors.  
A simple counting argument shows that the number of elements of the CSCO is
$N$ for the case of addition of $N$ arbitrary angular momenta.%
\footnote{The general theorem guarantees that the minimal number of elements
  of CSCO for finite dimensional representations can always be reduced to one. 
  In this paper, however, we adopt the direct sum decomposition (\ref{WSdual}) 
  and wish to find the CSCO according to this decomposition.} 
This follows from the fact that the direct product representation is given by
the joint eigenvalues of the $z$-components of the individual angular
momentum. 
To be precise, the states are labeled by $2N$ quantum numbers in the direct
product representation.  
Upon denoting the individual spin operator by
$\vec{J}_{\ell}=\vec{\sigma}_{\ell}/2$ ($\ell=1,2,\dots,N$),  
where the $\vec{\sigma}_{\ell}$s are the Pauli spin operators constituting the
Lie algebra $su(2)$,  
the squares are $J_{\ell}^{2}=\vec{J}_{\ell}\cdot \vec{J}_{\ell}$ and the
$z$-components of the individual angular momenta are $J_{\ell z}$.%
\footnote{The spin operators act on the full Hilbert space, as illustrated by
  ${\vec{\sigma}_2=I_2\otimes\vec{\sigma}\otimes I_2\otimes I_2\otimes\cdots
\otimes I_2}$.
} 
When considering systems in which each angular momentum is fixed, 
we will not write the eigenvalues of the Casimir operator 
$J_{\ell}^{2}$ explicitly. 
This paper deals with such a system and the total number $N$ of constituents 
thus determines the representation. 

From the general theory of quantum angular momentum, two of the commuting
operators in the CSCO are immediate. 
The first one is the Casimir operator of the rotational group, which labels
the total angular momentum quantum number, and the second is the $z$-component
of the total angular momentum. Therefore, the minimal number of elements in
the CSCO that are still to be constructed is $N-2$. 
These remaining $N-2$ operators are usually referred to as the missing label
operators (MLOs), and finding the MLOs has been a standard but rather difficult
problem in the representation theory~\cite{MLO1,MLO2,MLO3}.  
Note that the CSCO is not unique in general and to list all possible families
of CSCO seems an untrackable problem except for some special cases. 

With the total angular momentum $\vec{J}=\sum_{\ell=1}^{N}\vec{J}_{\ell}$, 
the above mentioned Casimir operator and the $z$-component of the total
angular momentum are $J^{2}=\vec{J}\cdot\vec{J}$ and 
$J_{z}=\sum_{\ell}J_{\ell z}$, respectively. 
The former operator $J^{2}$ specifies the angular momentum 
space $j$ in (\ref{decom}) and the partition $\nu$ in (\ref{WSdual}). 
The latter operator determines the representation $\mathcal{D}_j$ in
(\ref{decom}) and the representation $\mathcal{R}_{\nu}$ in (\ref{WSdual}). 
By definition, the MLOs commute with $J^{2}$ and $J_{z}$, 
and their eigenvalues are non-degenerate within each subspace specified by 
the common eigenvalues of $J^{2}$ and $J_{z}$. 
Without loss of generality, these non-degenerate eigenvalues can be chosen as
real, and thus the MLOs can be given by hermitian operators. 
It is not difficult to show that the MLOs live only in the subspace 
$\mathcal{S}_{\nu}$, and they are expressed as the superpositions of the
unitary permutation operators.  
This leads to the key observation that the MLO problem 
for $N$ spin-$j$ systems can be solved by finding a suitable unitary
representation of the symmetric group $S_{N}$.

\subsection{Symmetric coupling of three spin-1/2 systems}
The representation theory of the symmetric group has been much
studied~\cite{symbooks}. 
In the standard treatment, 
the irreducible representations can be constructed in real matrix
forms by employing the canonical subgroup chain  
\begin{equation} \label{stdchain}
S_{N}\supset S_{N-1} \supset \dots \supset S_{2}.
\end{equation} 

In the simplest case of $S_{3}$, for example, there exist three different 
irreducible representations corresponding to three possible Young tableaux. 
Besides trivial one-dimensional representations for the totally symmetric and 
anti-symmetric subspaces, the remaining non-trivial one is the 
two-dimensional subspace. 
In the above choice of subgroup chain (\ref{stdchain}), 
one can choose three possible proper subgroups $S_{2}(12)$, 
$S_{2}(23)$, and $S_{2}(31)$ where $S_2(i_{1},i_{2})$ is the transposition
subgroup between two indices $(i_{1}i_{2})$, e.g., 
$S_{2}(12)=\{P_{123},P_{213}\}$.  
When the reduction $S_{3}\supset S_{2}(12)$ is adopted, 
the representations corresponding to the decomposition (\ref{WSdual2}) are  
\begin{eqnArray} \label{j12}
P_{213}\ &\widehat{=}&\ds I_{1}\otimes I_{4}\oplus 
\left(\begin{array}{cc}1 & 0 \\0 & -1\end{array}\right)
\otimes I_{2}, \nonumber\\[4ex]
P_{132}\ &\widehat{=}&\ds I_{1}\otimes I_{4}\oplus\frac{1}{2}
\left(\begin{array}{cc}-1 & \sqrt{3} \\ \sqrt{3} & 1\end{array}\right)
\otimes I_{2},  
\end{eqnArray}
and the other representations for the elements of $S_{3}$ are generated by 
the combinations of these two transpositions. 
Since a particular proper subgroup $S_{2}(12)$ is diagonalized, 
the symmetry of the subspace representation labeled by the partition 
$\nu=(\nu_{1},\nu_{2})$ is determined by the subgroup $S_{2}(12)$, 
i.e., we have invariance under the transposition between the two indices 
$(1,2)$.  

With these observations, the coupling of three spin-1/2 systems is solved by 
specifying the MLOs. 
As discussed, the number of MLOs is $1$ for the $N=3$ case. 
This operator is identified with the transposition operator $P_{213}$ by
adopting the representation (\ref{j12}).  
Upon noting that the transposition operator between two given systems
$(k,\ell)$ is expressed in terms of the Pauli operators $\vec{\sigma}_{k}$  
and $\vec{\sigma}_{\ell}$ ($k\neq \ell$) by 
\begin{equation} \label{Pkl}
P_{k\ell}=\frac{1}{2}(I_{8}+\vec{\sigma}_{k}\cdot \vec{\sigma}_{\ell})=P_{\ell k}, 
\end{equation}
the MLO is written as $P_{213}=(I_8+\vec{\sigma}_{1}\cdot \vec{\sigma}_{2})/2$. 
It is straightforward to see the correspondence with the standard binary
coupling, in which the MLO is given by the square of the intermediate coupled
angular momenta.  
In the above choice of MLO $P_{213}$, the corresponding operator 
in terms of angular momenta is 
\begin{equation}
J_{12}^{2}=\vec{J}_{12}\cdot\vec{J}_{12}
=\frac{1}{2}(3I_8+\vec{\sigma}_{1}\cdot \vec{\sigma}_{2})=P_{123}+P_{213}, 
\end{equation}
where $\vec{J}_{jk}=\vec{J}_{k}+\vec{J}_{\ell}$ is the intermediate angular
momentum.   
Because the identity $P_{123}=I_{8}$ is irrelevant as far as MLOs are concerned, 
$P_{213}$ and $J_{12}^{2}$ are essentially the same MLO. 
This illustrates the claim that the coupling of identical angular momenta is
obtained by the representation theory of the symmetric group. 
The other possible MLOs are also found to be $J_{23}^{2}=P_{123}+P_{132}$ and 
$J_{31}^{2}=P_{123}+P_{321}$. 
Importantly, all three representations are related through the action of 
unitary transformations, and the subject matter of recoupling theory is to
study the relationships among these different representations. 

We now show a different coupling scheme by reconsidering the 
subgroup chain (\ref{stdchain}). 
It is well-known that the cyclic permutation is also a proper subgroup of
$S_{N}$.  
Denoting the $N$-cyclic permutation group by $C_{N}$, the alternative subgroup
chain is  
\begin{equation} \label{symchain}
S_{N}\supset C_{N} \supset C_{N-1}\supset \dots \supset C_{2}=S_{2}.
\end{equation} 
In the case of three spin-1/2 constituents, the cyclic group of order $3$ is 
$C_{3}=\{P_{123},P_{231},P_{312} \}$, and the natural representation of
$C_{3}$ is the diagonal matrix where the elements are powers of the basic 3rd
root of unity. 
Thus, two-dimensional representations for 
the elements of $S_{3}$ can be generated by the combination of the $3$-cycle 
permutation $P_{231}$ and $2$-cycle permutation $P_{213}$. 
They are 
\begin{eqnArray} \label{c3s12}
P_{231} &\widehat{=}&\ds I_{1}\otimes I_{4}\oplus
\left(\begin{array}{cc}\omega_{3} & 0 \\0 & \omega_{3}^{2}\end{array}\right)
\otimes I_{2}, \nonumber\\[4ex]
P_{213} &\widehat{=}&\ds I_{1}\otimes I_{4}\oplus
\left(\begin{array}{cc}0 & 1 \\ 1 & 0\end{array}\right)
\otimes I_{2},  
\end{eqnArray}
where $\omega_{d}=\exp(2\pi \I/d)$ is the basic $d$th root of unity. 
One can, of course, check that the two representations in (\ref{j12}) and
(\ref{c3s12}) are related by the unitary matrix 
\begin{equation}
 I_{1}\otimes I_{4}\oplus\frac{1}{\sqrt{2}}
\left(\begin{array}{cc}1 & 1 \\ -\I & \I\end{array}\right)\otimes I_{2},  
\end{equation}
but this representation has higher symmetry than the previous one. 
This is because the cyclic permutation subgroup $C_3$ has order $3$ rather
than $2$ for the transposition subgroup.  
We remark that this choice of irreducible representation is only possible with
complex numbers.  

The MLOs are readily found through the relation $P_{231}=P_{132}P_{213}$ and the
Pauli operator representation of the transition operator (\ref{Pkl}), 
\begin{equation}
P_{231}=\frac{1}{4}\bigl[ I_{8}+ \vec{\sigma}_{1}\cdot \vec{\sigma}_{2}
+\vec{\sigma}_{2}\cdot \vec{\sigma}_{3}
+\I \vec{\sigma}_{1}\cdot (\vec{\sigma}_{2}\times\vec{\sigma}_{3}) \bigr] .
\end{equation}
Since the other element of the cyclic permutation operator $P_{312}$ is
equally good for the MLOs, the two cyclic permutations can be combined to give 
the real one of the MLOs as 
\begin{eqnArray}
K&\equiv&\ds\frac{-\I}{\sqrt{3}}(P_{231}-P_{312})= 
\frac{1}{\sqrt{12}}\vec{\sigma}_{1}\cdot 
(\vec{\sigma}_{2}\times\vec{\sigma}_{3}) \nonumber\\[3ex]
&\widehat{=}& \ds \mathbf{0}_1\otimes I_{4}\oplus
\left(\begin{array}{cc} 1& 0 \\ 0 &-1 \end{array}\right)
\otimes I_{2}.
\end{eqnArray}
This form of the MLO is more convenient for us for two reasons. 
Firstly, the eigenvalues of $K$ are $\pm 1$ and, secondly, $K$ is
projected onto the two-dimensional subspace $\nu=(2,1)$.  
The corresponding angular momentum states in the $j=1/2$ subspace are 
\begin{eqnArray}
|1/2,1/2;\lambda\rangle&=&
\ds\frac{1}{\sqrt{3}}\bigl(|100\rangle \omega_3^{\lambda}
+|010\rangle\omega_3^{2\lambda} + |001\rangle\bigr),\\[2.5ex]
|1/2,-1/2;\lambda\rangle&=&
\ds-\frac{1}{\sqrt{3}}\bigl(|011\rangle \omega_3^{\lambda}
+|101\rangle\omega_3^{2\lambda} + |110\rangle\bigr), 
\end{eqnArray}
where ${\lambda=1,2}$ and we denote by $|0\rangle$ and $|1\rangle$ the kets
with $m=1/2$ and $m=-1/2$, respectively, for the states of the single spin-1/2
constituents.    

Note that this CSCO $\{J^{2},J_{z},K \}$ is indeed identical with that of
\cite{sym1,sym2},  
where they proved that this set is the CSCO for the general coupling of three
angular momenta $\vec{J}_{\ell}$ ($\ell=1,2,3$) with the MLO 
$K=\vec{J}_{1}\cdot (\vec{J}_{2}\times\vec{J}_{3})$. 
Although this novel coupling scheme was discovered almost half a century ago, 
the generalization to the case of more than three angular momenta is 
--- to our knowledge --- still an open problem. 
The reason seems that there exist infinitely many different MLOs, which are 
equivalent in general. 
It is then rather hard to conclude that a particular choice of MLOs provides the
symmetric coupling without a precise definition of the symmetric coupling.  
In the next subsection, we attempt to give a possible definition of the
symmetric coupling in the case of identical spin systems.

\subsection{A proposal for symmetric coupling}
We define the symmetric coupling of identical angular momenta, not necessarily
for the case of spin-1/2 systems, as follows:
\begin{equation}\label{eq:def-sc}
\hspace*{-0.1\textwidth}\parbox{0.8\textwidth}{%
If the MLO projected onto a subspace labeled by 
the partition $\nu$ can be chosen as a linear combination of cyclic
permutation operators with the possible maximal order in this subspace, then
identical angular momenta are  
said to be coupled \textit{symmetrically} within the subspace. 
} 
\end{equation}
The word ``symmetric coupling'' refers to the fact that this coupling respects 
the cyclic permutation symmetry within each subspace. 
In this definition, the possible maximal order of the cyclic permutation
operator still needs to be stated explicitly. 
In this paper, we set the order of the cyclic permutation
subgroup to be $s(\nu)+1$ for the partition $\nu$. 
If the order is less than $s(\nu)+1$, we do not have a symmetric coupling in
the subspace.  
Note that the cyclic permutation group is abelian and hence all linear
combinations are equally good as the MLOs as long as they have non-degenerate
eigenvalues.   
Another important consequence of the abelian property is that the obtained 
irreducible basis is invariant under the same cyclic permutation.  

We emphasize that this definition is rather limited since it only applies to 
the case of addition of identical angular momenta for a few angular momentum
systems.  
In particular, the number $s(\nu)+1$ becomes greater than $N$ for 
more than four spin-1/2 systems and we need to refine the meaning of 
the possible maximal order of the cyclic permutation operator properly. 
According to the definition (\ref{eq:def-sc}), for example, it follows that
five spin-1/2 systems cannot be coupled symmetrically in the subspace
$\nu=(3,2)$ ($j=1/2$) whose degeneracy is $5$.  
Nevertheless, we will show that a symmetric coupling of four spin-1/2 systems
is possible.  
The coupling of more spin-1/2 systems and other more general cases are left to
future studies.

\subsection{Relation to reference-frame-free subsystems}
Owing to the Weyl-Schur duality, the role of the permutation symmetry is clear
in the tensor-product Hilbert space. 
This leads to the idea of constructing a reference-frame-free (RFF) 
subsystem, or a RFF qudit, in the context of quantum information theory. 
A RFF qudit is a $d$-dimensional subsystem of a composite system, which
remains invariant under the same unitary transformation on the constituents of
the composite system.  
It is called a rotationally invariant qudit when considering the invariant
subsystems under the collective rotation rather than the general unitary
transformations.  
The RFF subsystems have many applications in quantum information 
and quantum computing~\cite{BRS}. 
The Werner state for two parties is the simplest example~\cite{werner}, 
and the generalization to more than two parties have been studied in this
context~\cite{EW, Egg, CK}. 

When dealing with spin-1/2 systems, any unitary transformation is equivalent 
to a rotation ($SU(2)\cong SO(3)$). 
Then, RFF subsystems can be described by a non-negative unit-trace density
operator in the subsystems, which commutes with the $N$-fold tensor product of
the rotation $u_{j}=\exp(\I \vec{n}\cdot\vec{J}_{j})$. 
Denoting the density operator by $\rho$ ($\rho \ge 0$, $\tr\{\rho\}=1$), 
the condition for the RFF subsystem reads 
\begin{equation}
{\left[\rho,\prod_{j=1}^{N}u_{j}\right]}
={\left[\rho,\Exp{\I \vec{n}\cdot\vec{J}}\right]}=0, 
\end{equation}
where $\vec{J}$ is the total angular momentum operator. 
From the properties of the Weyl-Schur duality, 
it immediately follows that the only possible form of the density operator for
the RFF subsystem is  
\begin{equation}
\rho^{\mathrm{rff}}=\bigoplus _{\nu\in\Par{N,2}} \rho_{s(\nu)}\otimes I_{r(\nu)}, 
\end{equation}
which is a linear combination of permutation operators $P_{i_1 i_2 \dots i_N}$. 
Therefore, the construction of RFF subsystems is essentially a problem of 
revealing the algebraic relationship between the permutation operators 
in the various subspaces, that is, to analyze the different choices of MLOs. 
In fact, the following stronger statement holds: 
\begin{equation}
\hspace*{-0.1\textwidth}\parbox{0.8\textwidth}{%
All possible MLOs for the subspace $\nu$ are linear combinations
of a non-degenerate RFF density operator and the projector  
onto the subspace~$\nu$, 
$$\alpha \rho_{\nu}^{{\mathrm{rff}}}+\beta I_{s(\nu)}, $$
where $\alpha$ and $\beta$ are coefficients.
} 
\end{equation}
In the Lie algebra theory, the algebra formed by RFF states is known as 
the (universal) enveloping algebra of $su(2)$~\cite{Jacob}. 

There are many ways to describe a quantum state defined in 
the $d$-dimensional Hilbert space $\mathbb{C}^d$. 
The simplest one is to use the $d$ orthonormal kets $|k\rangle$
$(k=1,2,\dots,d)$ that are orthonormal and complete,
\begin{equation}
\quad \langle k|k'\rangle=\delta_{kk'}, 
\quad \sum_{k=1}^{d}|k\rangle\langle k|=I_{d}, 
\end{equation}
and thus form a basis in $\mathbb{C}^d$. 
Correspondingly, the $d^2$ operators
\begin{equation} 
Q_{k\ell}\equiv |k\rangle\langle \ell | \quad (k,\ell=1,2,\dots,d) 
\end{equation}
that satisfy the closure relation $Q_{k\ell}Q_{k'\ell'}=\delta_{\ell k'}Q_{k\ell'}$, 
form a basis for the $d$-dimensional operator-algebra space, i.e., 
the $d$-dimensional matrix ring over $\mathbb{C}$. 
We call these $d^2$ operators the RFF basis operators. 
The state of a $d$-dimensional quantum system can be represented as a
unit-trace $d\times d$ matrix that is semi-definite positive.
Another possible way is to expand the state in terms of the generators for a
$SU(d)$ Lie group with real coefficients. 
The standard Gell-Mann matrices together with the semi-definite 
positivity requirement provide a proper $d$-level quantum
state~\cite{kimura,bryd}.  
The third option is to use the unitary Heisenberg-Weyl-Schwinger (HWS)
operator basis~\cite{schwinger}.  
The complete set of unitary operators is given by 
$U^{k}V^{\ell}$ ($k,\ell=1,2,\dots,d$), 
where the unitary operators $U$ and $V$ have the period $d$, i.e., $U^d=V^d=1$, 
and satisfy the Weyl commutation relation 
$U^{k}V^{\ell}=\omega_d^{-k\ell}V^{\ell}U^{k}$ 
with $\omega_{d}=\exp(2\pi \I/d)$ as above. 
From the mathematical point of view, all three 
operator bases and any other ones are equivalent in the sense that we can
convert one description into the others through a bijective mapping, 
and an advantage over others may show up depending upon the problem of
interest.  
In the following, the first method is mainly considered, 
and the basis operators $Q_{k\ell}^{\nu}$ are to be constructed. 

The Weyl-Schur duality and the decomposition in (\ref{WSdual3}) already provide 
the relation between any permutation operator and the basis operators
$Q_{k\ell}^{\nu}$ in the subspace labeled by a partition $\nu$. 
Firstly, define the RFF basis operators 
in the subspace $\nu$ by
\begin{equation}
Q_{k\ell}^{\nu}=|k\rangle\langle \ell |\otimes I_{r(\nu)}, 
\end{equation}
where the operator $|k\rangle\langle \ell |$ lives in the subspace 
${\cal S}_{\nu}$. 
Then, they form the RFF basis operators satisfying 
$Q^{\nu}_{k\ell}Q^{\nu'}_{k'\ell'}=\delta_{\nu\nu'}\delta_{\ell k'}Q_{k\ell'}^{\nu}$, and 
the permutation operator can be written as  
\begin{equation} \label{StoRFF}
P_{i_{1}i_{2}\dots i_{N}}=\sum _{\nu\in\Par{N,2}} 
\sum_{k,\ell=1}^{s(\nu)}p_{k\ell}^{\nu}\,Q_{k\ell}^{\nu}.  
\end{equation}
Here, the coefficients $p_{k\ell}^{\nu}$ are elements of the matrix
representation of the symmetric group within the subspace $\nu$. 
Using the RFF basis operators, the MLO $M_{\nu}$ can be expressed as a linear
combination of diagonal elements of them with different coefficients,  
\begin{equation} 
M_{\nu}=\sum_{\lambda=1}^{s(\nu)}q_{\lambda}^{\nu} Q_{\lambda\lambda}^{\nu}. 
\end{equation}

In the example of three spin-1/2 systems, the relation (\ref{StoRFF}) for 
the subgroup chain (\ref{symchain}) reads 
\begin{eqnArray}\nonumber
P_{123}&=&Q_{11}^{(3,0)}+Q_{11}^{(2,1)}+Q_{22}^{(2,1)},\\[1.5ex] \nonumber
P_{231}&=&Q_{11}^{(3,0)}+\omega_{3}Q_{11}^{(2,1)}+\omega_{3}^{2}Q_{22}^{(2,1)}
=(P_{312})^{\dagger}, \\[1.5ex] \nonumber
P_{213}&=&Q_{11}^{(3,0)}+Q_{12}^{(2,1)}+Q_{21}^{(2,1)},\\[1.5ex] \label{eqS3}
P_{132}&=&Q_{11}^{(3,0)}+\omega_{3}Q_{12}^{(2,1)}+\omega_{3}^{2}Q_{21}^{(2,1)},
\\[1.5ex]  
P_{321}&=&Q_{11}^{(3,0)}+\omega_{3}^{2}Q_{12}^{(2,1)}+\omega_{3}Q_{21}^{(2,1)}. 
\end{eqnArray}
Although these equations seem to be over-determined at first sight, i.e., six
equations for five variables $Q_{k\ell}^{\nu}$, only five equations are
actually linearly independent. 
Note that in general there are $N!$ linear equations for the RFF basis
operators of which the total number is 
\mbox{\small$\displaystyle\sum_{\nu\in\Par{N,2}}\hspace{-1em}s(\nu)^2$}. 
This number is always less than $N!$ and, therefore, the occurrence of
linearly dependent equations is generic. 
This is due to the fact that the entire representation 
space is not exhausted when considering a problem of $N$ spin-1/2 systems. 
The complete set of representations can be obtained for the case of $N$ 
spin-$j$ ($j=N/2+1$) systems, where the following relation holds:  
\begin{equation}
\sum_{\nu\in\Par{N,N}}s(\nu)^2=N! . 
\end{equation}

By converting equations (\ref{eqS3}), we obtain the operators
$Q_{k\ell}^{\nu}$ in terms of the permutation operators:  
\begin{eqnArray}
\left(\begin{array}{c} 
Q_{11}^{(3,0)}\\ Q_{11}^{(2,1)}\\  Q_{22}^{(2,1)}\end{array}\right)
&=&\ds \frac{1}{3}\left(\begin{array}{ccc} 
1&1&1\\ 1&\omega_3^2&\omega_3\\ 1&\omega_3&\omega_3^2
\end{array}\right)
\left(\begin{array}{c} P_{123}\\ P_{231}\\  P_{312}\end{array}\right),\\[4ex]
\hspace*{-2em}
\left(\begin{array}{c} 
Q_{11}^{(3,0)}\\ Q_{12}^{(2,1)}\\  Q_{21}^{(2,1)}\end{array}\right)
&=&\ds \frac{1}{3}\left(\begin{array}{ccc} 
1&1&1\\ 1&\omega_3^2&\omega_3\\ 1&\omega_3&\omega_3^2
\end{array}\right)
\left(\begin{array}{c} P_{213}\\ P_{132}\\  P_{321}\end{array}\right). 
\end{eqnArray}
In terms of the Pauli spin operators, they read 
\begin{eqnArray}
Q_{11}^{(3,0)}
&=&\ds\frac{1}{2}I_{8}+\frac{1}{6}\left(\vec{\sigma}_{1}\cdot \vec{\sigma}_{2}
+\vec{\sigma}_{2}\cdot \vec{\sigma}_{3}
+\vec{\sigma}_{3}\cdot \vec{\sigma}_{1} \right),\\[2ex]
\hspace*{-2em}\left.\begin{array}{r}
Q_{11}^{(2,1)}\\[1ex]
Q_{22}^{(2,1)}
\end{array}\right\}
&=&\ds\frac{1}{4}I_{8}-\frac{1}{12}\left(\vec{\sigma}_{1}\cdot \vec{\sigma}_{2}
+\vec{\sigma}_{2}\cdot \vec{\sigma}_{3}
+\vec{\sigma}_{3}\cdot \vec{\sigma}_{1} \right)\\[2ex]
&&\ds\mbox{}\pm\frac{1}{\sqrt{48}}\vec{\sigma}_{1}
\cdot (\vec{\sigma}_{2}\times\vec{\sigma}_{3}),\\[2ex]
Q_{12}^{(2,1)}&=&\ds\left(Q_{21}^{(2,1)}\right)^{\dagger}
=\ds\frac{1}{6}\left(\vec{\sigma}_{1}\cdot \vec{\sigma}_{2}
+\omega_{3}^{2}\vec{\sigma}_{2}\cdot \vec{\sigma}_{3}
+\omega_{3}\vec{\sigma}_{3}\cdot \vec{\sigma}_{1} \right). 
\end{eqnArray}
With these basis operators $Q_{11}^{(3,0)}$ and $Q_{k\ell}^{(2,1)}$, we can express 
any RFF state as a linear combination of them, and this completes the analysis
of three spin-1/2 systems.

\section{Symmetric coupling for the second-largest angular momentum}
\label{sec:3}
In this section, we investigate a symmetric coupling of $N$ spin-1/2 systems 
within the second-largest angular momentum subspace~\cite{JGB}. 
Since the second-largest angular momentum has $N-1$ components and 
its multiplicity is $c_{N/2-1}=N-1$, there are $(N-1)^2$ states in the
subsystem in total.  
The basic ingredient is the grading of the $N$ spin-1/2 constituents 
$\vec{J}_{\ell}=\vec{\sigma}_{\ell}/2$ ($\ell=1,2,\dots,N$), 
\begin{equation} \label{grading}
\vec{\Sigma}(\lambda)=\sum_{\ell=1}^{N}\omega_{N}^{\lambda \ell}\vec{J}_{\ell}. 	
\end{equation}
The parameter $\lambda$ takes the values $\lambda=1,2,\dots,N$ with modulo $N$
and the $\lambda=N$ case reduces to the usual total angular momentum operator  
$\vec{\Sigma}(N)=\vec{J}=\sum_{\ell=1}^{N}\vec{\sigma}_{\ell}/2$.  
They satisfy ${\vec{\Sigma}(\lambda)}^{\dagger}=\vec{\Sigma}(N-\lambda)$ and
the commutation relation  
\begin{equation}
\Bigl[\vec{a}\cdot\vec{\Sigma}(\lambda),\vec{b}\cdot\vec{\Sigma}(\lambda')\Bigr]
=\I (\vec{a}\times\vec{b})\cdot\vec{\Sigma} (\lambda+\lambda') 
\end{equation}
for all numerical vectors $\vec{a}$ and $\vec{b}$.
With this commutation relation, the $\vec{\Sigma}(\lambda)$s form a so-called
graded Lie algebra labeled by the integers $\lambda~(\mathrm{mod}\ N)$. 

Denoting the kets for the single spin-1/2 states with $m=1/2$ and $m=-1/2$ by
$|0\rangle$ and $|1\rangle$ as before, the ket for the state with maximal
values of both $j$ and $m$, i.e., $j_1=m_1=N/2$, is  
\begin{equation}
|0_{N}\rangle\equiv|0\rangle^{\otimes N},
\end{equation}
and successive applications of the lowering operator $J_-=J_x-\I J_y$ yield 
all states for the maximal angular momentum space, 
\begin{equation}
|j_1,m_1\rangle=\sqrt{\frac{(j_1+m_1)!}{j_1!\,(j_1-m_1)!}}
\;(J_{-})^{j_1-m_1}|0_{N}\rangle .
\end{equation}
The highest states with $j_2=m_{2}=N/2-1$ for the second-largest angular
momentum space are given by the action of $N-1$ lowering operators 
\begin{equation} \label{omega}
\Sigma_{-}(\lambda)=\Sigma_{x}(\lambda)-\I\Sigma_{y}(\lambda)
=\sum_{\ell=1}^{N}\omega_{N}^{\lambda \ell}J_{\ell -} .
\end{equation}
onto the highest state $|0_{N}\rangle$ in the largest angular momentum space as 
\begin{equation}
|j_2,j_2;\lambda\rangle= \frac{1}{\sqrt{N}}\Sigma_{-}(\lambda)|0_{N}\rangle 
\quad (\lambda=1,2,\dots,N-1),
\end{equation}
and successive applications of $J_{-}$ give all the remaining states 
$|j_2,m_2;\lambda\rangle$ with $m_2=-j_2,-j_2+1,\dots,j_2$.
Since $\Sigma_{-}$ and $J_{-}$ commute with each other, the resulting states are 
\begin{eqnArray} \label{symmcoup}\nonumber
|j_2,m_2;\lambda\rangle &=&\ds\sqrt{\frac{(j_2+m_2)!}{N(2 j_2)!(j_2-m_2)!}}
\;\Sigma_-(\lambda)J_{-}^{j_2-m_2}|0_{N}\rangle \\[4ex] 
& =&\ds\sqrt{\frac{2j_1-1}{(j_1+m_2+1)(j_1+m_2)}}\;
\Sigma_-(\lambda) |j_1,m_2+1\rangle ,
\end{eqnArray}
which are orthonormal in the $j_2$ subsystem, 
\begin{eqnArray}
&&\ds\langle j_2,m_2^{\phantom{\ }};\lambda|j_2,m_2';\lambda'\rangle
=\delta_{m_2^{\phantom{\ }}m_2'}\delta_{\lambda\lambda'},\\[2ex] 
&& \ds\sum_{\lambda=1}^{N-1}\sum_{m_2=-j_2}^{j_2}
|j_2,m_2;\lambda\rangle\langle j_2,m_2;\lambda|=I_{j_2}.
\end{eqnArray}
Here, $I_{j_2}$ is the projector onto the angular momentum $j_2$ subsystem.
This projector is a polynomial of the Casimir operator 
${J^2=\vec{J}\cdot\vec{J}}$,
\begin{equation}
I_{j_2}=\prod_{j\in J, j\neq j_2} \frac{J^2-j(j+1)}
                                   {j_2(j_2+1)-j(j+1)}.
\end{equation} 
Their orthogonality is more transparent when the states (\ref{symmcoup}) are
written as  
\begin{equation}
|j_2,m_2;\lambda\rangle 
= \frac{1}{\sqrt{N}}\sum_{\ell=1}^{N}|\ell;j_2,m_2\rangle\omega_{N}^{\lambda \ell},
\end{equation}
where the $N$ states $|\ell;j_2,m_2\rangle$ form a pyramid, 
\begin{equation}
\langle \ell;j_2,m_2|\ell';j_2,m_2'\rangle
=\delta_{m_2m'_2}\left(\delta_{\ell\ell'}+\frac{j_2-m_2}{j_2+m_2+1}\right).
\end{equation}
From the Weyl-Schur duality, the second-largest angular momentum states 
$|j_2,m_2;\lambda\rangle$ correspond to the partition 
$\nu=(N-1,1)$ in the decomposition (\ref{WSdual}), and they are 
$(N-1)$-dimensional irreducible representations of the symmetric group $S_N$. 

We note that the discrete Fourier transformation that we chose in
(\ref{omega}) is just one of many possibilities for defining the
$\Sigma_{-}(\lambda)$s and thus the kets $|j_2,m_2;\lambda\rangle$. 
More generally, any unitary $(N-1)\times (N-1)$ matrix, 
with $N$th-row matrix elements $U_{N\ell}=1$, can serve in 
$\Sigma_{-}(\lambda)=\sum_{\ell=1}^{N} U_{\lambda \ell}\sigma_{-}^{(\ell)}$. 
For the specific choice of the discrete Fourier matrix, the projectors 
$|j_2,m_2;\lambda\rangle\langle j_2,m_2;\lambda|$ are invariant under the
cyclic permutation subgroup of order $N$ generated by $P_{23\cdots N1}$.  

Following the construction for the $N=3$ case, it is natural to look for the
representation of $S_{N}$ that possesses a cyclic permutation symmetry of
order $N$.  
In fact, without exploring the representation theory, 
we can immediately construct the $(N-1)^{2}$ RFF basis operators from the
states (\ref{symmcoup}) by tracing over the quantum number $m_{2}$,  
\begin{equation}
Q_{\lambda\lambda'}^{(N-1,1)}=\sum_{m_2=-j_2}^{j_2}|j_2,m_2,\lambda\rangle
\langle j_2,m_2,\lambda' |
=\left(Q_{\lambda'\lambda}^{(N-1,1)}\right)^{\dagger} .
\end{equation}
Indeed, we can check the properties
\begin{eqnArray} \label{}
\ds{\left[\vec{J}, Q_{\lambda\lambda'}^{(N-1,1)}\right]}&=&0, \\[2ex]
Q_{\lambda\lambda'}^{(N-1,1)}Q_{\lambda''\lambda'''}^{(N-1,1)}
&=&\delta_{\lambda'\lambda''}Q_{\lambda\lambda'''}^{(N-1,1)}. 
\end{eqnArray} 

The explicit construction in (\ref{symmcoup}) of the angular momentum states
in the subspace with $j_2=N/2-1$, enables us to express the MLO for the
second-largest angular momentum subspace as a linear combination of diagonal
elements of the RFF basis operators with different coefficients.  
It is then straightforward but rather tedious to rewrite it in terms of
individual Pauli spin operators.  
In the next subsection, we provide an explicit construction of the MLO for the
second-largest angular momentum subspace in terms of cyclic permutation
operators.

\subsection{Missing label operator for the second-largest angular momentum
  subspace} \label{sec:3-1}
In the construction of the second-largest angular momentum states in
(\ref{symmcoup}), the cyclic permutation of order $N$ is respected according
to the subgroup chain of the  permutation group, i.e., $S_N\supset C_N$. 
Utilizing this fact we now construct the MLO for 
the second-largest angular momentum subspace in terms of the Pauli operators. 
In the following, we restrict ourselves to representations of the permutation
group elements  
within the second-largest angular momentum subspace whose dimension is $N-1$. 

The cyclic permutation operator $C$ that transforms the index
$(1,2,\dots,N-1,N)$ to $(2,3,\dots,N,1)$ is given by 
\begin{equation}
C=P_{23\cdots N1}=P_{N\,N-1}P_{N-1\,N-2}\dots P_{32}P_{21},
\end{equation}
where, as before, $P_{ij}$ denotes the transposition operator between the two
indices $(i,j)$,
\begin{equation}
P_{ij}=\frac{1}{2}\bigl(I_{2^N}+\vec{\sigma}_i\cdot\vec{\sigma}_j\bigr).
\end{equation}
As stated, we choose a diagonal representation of the above cyclic permutation
within the second-largest angular momentum subspace as  
\begin{equation}
C \mathrel{\dot{=}_{j_2}}
\left(\begin{array}{cccc}
\omega_N & & & \\& \omega_N^2 & & \mbox{\huge{0}} \\ 
 \mbox{\ \huge{0}} & & \ddots & \\
& & & \omega_N^{N-1}
 \end{array}\right)\otimes I_{N-1}
\end{equation}
where ``$\,\dot{=}_{j_2}\,$'' indicates the restriction to the $j_2$ subspace.
The symmetrically constructed second-largest angular momentum states are 
the eigenstates of cyclic permutation operator 
\begin{equation}
C|j_2,m_2;\lambda\rangle=|j_2,m_2;\lambda\rangle\omega_N^{\lambda}.  
\end{equation}
This is a consequence of the permutation invariance for the largest angular
momentum states and the commutation relation $C\Sigma_-(\lambda)= 
\omega_N^{\lambda}\Sigma_-(\lambda) C$.  

It is straightforward to construct  $N$ orthogonal projectors by the inverse Fourier
transform of the 
permutation operators $\{C,C^2,\dots, C^{N-1},C^N=I_{2^N}\}$,
\begin{equation}
P_c(\lambda)=\frac{1}{N}\sum_{k=1}^N\omega_N^{-k\lambda}C^k. 
\end{equation}
They have the following representation within the second-largest angular
momentum subspace  
\begin{equation}
P_c(\lambda) \mathrel{\dot{=}_{j_2}} 
\left(\begin{array}{cccc}
\delta_{\lambda\,1} & & & \\& \delta_{\lambda\,2} & & 
\mbox{\huge{0}} \\ \mbox{\ \huge{0}} & & \ddots & \\
& & & \delta_{\lambda\,N-1}
 \end{array}\right)\otimes I_{N-1},  
\end{equation}
and satisfy the orthogonal relation 
\begin{equation}
P_c(\lambda)P_c(\lambda')=\delta_{\lambda\,\lambda'}P_c(\lambda). 
\end{equation}
We thus obtain  
\begin{equation}
P_c(\lambda)|j_2,m_2;\lambda'\rangle
=|j_2,m_2;\lambda\rangle\delta_{\lambda\,\lambda'}.
\end{equation}
These relations are obtained directly from the commutation relation 
$P_c(\lambda) \Sigma_-(\lambda')=\Sigma_-(\lambda')P_c(\lambda-\lambda')$ 
and the observation that 
$P_c(\lambda)|j_1,m_1\rangle=|j_1,m_1\rangle\delta_{\lambda N}$.
We remark that the rank of the projectors 
$P_c(\lambda)$ ($\lambda=1,2,\dots,N-1$) is 
greater than $N-1$ in general. 
It follows that they project onto not only the second-largest angular momentum
subspace but other lower angular momentum subspaces as well.  

With the above result, we can express the MLO for the
second-largest angular momentum subspace as  
\begin{equation}
M_{j_2}=\sum_{\lambda=1}^{N-1} q_{\lambda}^{(N-1,1)} P_c(\lambda)
\end{equation}
with $N-1$ different coefficients $q_{\lambda}^{(N-1,1)}$. 
The choice  $q_{\lambda}^{(N-1,1)}=j_2+1-\lambda$ reads 
\begin{eqnArray}
M_{j_2}&=&\ds\sum_{k=1}^{N-1} \gamma_k C^k,\\[3.5ex]\label{eq:52}
\gamma_k&=&\ds \frac{1}{N}\sum_{\lambda=1}^{N-1}\Big(\frac{N}{2}-\lambda\Big)\omega_N^{k\lambda}.
\end{eqnArray} 
The coefficients $\gamma_k$ have the properties $\gamma_N=0$ and $\gamma_{N-k}=-\gamma_k$, 
and the MLO is stated as the summation of $C^k-C^{N-k}$ with certain
coefficients.

\section{Symmetric coupling for four spin-1/2 systems}\label{sec:4}
In this section the symmetric coupling for four spin-1/2 systems is
accomplished along the ideas described in section~\ref{sec:2}. 
The standard binary coupling of four angular momenta is known under names such
as the L-S coupling and the j-j coupling~\cite{AMinQP1, AMinQP2}.  
In the first step, two angular momenta $\vec{J}_{1}$ and $\vec{J}_{2}$ are
coupled and similarly the remaining two $\vec{J}_{3}$ and $\vec{J}_{4}$ are
coupled. 
Then, the newly coupled angular momenta  $\vec{J}_{12}$ and $\vec{J}_{34}$ are
coupled in the last stage.  
With this particular choice of intermediate angular momentum states, the CSCO
are 
\begin{equation}
\mathrm{CSCO}_{12|34}=\{ J^{2},J_{z},J_{12}^{2},J_{34}^{2} \}.   
\end{equation}
Since there are three inequivalent ways of paring four angular momenta as the
intermediate states, there are two more choices for CSCO, namely
$\mathrm{CSCO}_{13|24}$  and $\mathrm{CSCO}_{14|23}$.

From the Weyl-Schur duality, the partition of four into two is 
$\Par{4,2}=\{(4,0)$, $(3,1)$, $(2,2)\}$, 
and the corresponding dimensions of the subspaces for the
symmetric group are $s(\nu)=1,3,2$, respectively.  
The total number of basis operators to span the RFF subsystems is thus
$1^{2}+3^{2}+2^{2}=14$.  
As noted earlier, this number is smaller than the number of elements in
$S_{4}$, namely $14<4!=24$.  
The cyclic permutation subgroup of order $4$, 
\begin{equation}\label{c4}
C_{4}=\{P_{1234},P_{2341}, P_{3412},P_{4123}\},
\end{equation}
is used to diagonalize the subspace $(3,1)$ ($j=1$), 
and similarly the cyclic permutation subgroup of order $3$
\begin{equation}
C_{3}(123)=\{P_{1234},P_{2314},P_{3124} \},
\end{equation}
is used for the subspace $(2,2)$ ($j=0$). We remark that there are 
two other cyclic permutation subgroup of order $4$ and three others of order
$3$.  
However, all other choices lead essentially to the same result. 
Following the subgroup chain (\ref{symchain}) together with the last
transposition permutation subgroup $S_{2}(12)$, 
we obtain the following unitary representations 
for the permutation operators: 
\begin{eqnArray} \label{S4cyc}
\hspace*{-2em}
P_{2341}\ &\widehat{=}&I_{1}\otimes I_{5}\oplus
\left(\begin{array}{ccc}\omega_{4} & 0 & 0 \\
0 & \omega_{4}^{2} & 0 \\0 & 0 & \omega_{4}^{3}
\end{array}\right)
\otimes I_{3}\oplus
\left(\begin{array}{cc}0&\omega_{3}^{2} \\ 
    \omega_{3}&0\end{array}\right)\otimes I_{1}, 
\\[4ex] \nonumber
\hspace*{-2em}
P_{2314}\ &\widehat{=}&I_{1}\otimes I_{5}\oplus\frac12 
\left(\begin{array}{ccc} \I & 1-\I &1 \\ 
1-\I &0 &1+\I \\ 
1 & 1+\I &-\I 
\end{array}\right)
\otimes I_{3}\oplus
\left(\begin{array}{cc}\omega_{3}&0 \\ 
0& \omega_{3}^{2}\end{array}\right)\otimes I_{1}, \\[4ex] \nonumber
\hspace*{-2em}
P_{2134}\ &\widehat{=}&I_{1}\otimes I_{5}\oplus\frac12
\left(\begin{array}{ccc} 1 & 1-\I &\I \\ 
1+\I &0 &1-\I \\ 
-\I & 1+\I &1 
\end{array}\right)
\otimes I_{3}\oplus
\left(\begin{array}{cc}0&1 \\ 
1&0\end{array}\right)\otimes I_{1}, 
\end{eqnArray}
and all other elements of the permutation group $S_{4}$ can be generated with
these three elements. 
Using the above choice of the representations, the basis operators 
$Q_{k\ell}^{\nu}$ can be expressed in terms of the permutation operators in a 
straightforward, if somewhat tedious, manner.
For completeness, we list all relations using the notations where 
$I=P_{1234}$, $P_{ij}$ for the transposition permutations, and  
\begin{eqnArray}\nonumber
C_1&=&P_{2341},\, C_1^2=P_{3412},\,C_1^3=P_{4123},\\[1.5ex] 
C_2&=&P_{2413},\, C_2^2=P_{4321},\, C_2^3=P_{3142},\\[1.5ex]
C_3&=&P_{3421},\, C_3^2=P_{2143},\, C_3^3=P_{4312},\\[1.5ex]
D_1&=&P_{1342},\,D_1^2=P_{1423},\\[1.5ex]
D_2&=&P_{3241},\,D_2^2=P_{4213},\\[1.5ex]
D_3&=&P_{2431},\,D_3^2=P_{4132},\\[1.5ex]
D_4&=&P_{2314},\,D_4^2=P_{3124}, 
\end{eqnArray}
for cyclic permutations as follows. 
For the subspaces with $\nu=(4,0)$ and $\nu=(3,1)$, we have two different
representations in terms of odd and even permutation operators as  
\begin{eqnArray}\nonumber\hspace*{-4em}
Q_{11}^{(4,0)}&=&\ds \frac{1}{12} 
\Bigl[I+\sum_{k=1}^3C_k^2+\sum_{l=1}^4(D_l+D_l^2)\Bigr],\\[1.5ex]
&=&\ds\frac{1}{12}\Bigl[ \sum_{k=1}^3(C_k+C_k^3)+\sum_{i>j}P_{ij}\Bigr],
\\[1.5ex]\hspace*{-4em}
\left.\begin{array}{r}
Q_{11}^{(3,1)}
\\[1ex]\hspace*{-4em}
Q_{33}^{(3,1)}
\end{array}\right\}&=&\ds
\frac{1}{4}(I-C_1^2)\mp\frac{\I}{4}\sum_{l=1}^4(D_l-D_l^2),
\\[1.5ex]\hspace*{-4em}
&=&\ds
\frac{1}{8}(P_{12}+P_{23}+P_{34}+P_{41})
\\[2ex]\hspace*{-4em}
&&\ds\mbox{}-\frac{1}{8}(C_2+C_2^3+C_3+C_3^3)
\mbox{}\mp\frac{\I}{4}(C_1-C_1^3)  ,
\\[1.5ex]\hspace*{-4em}
Q_{22}^{(3,1)}&=&\ds\frac{1}{4}(I+C_1^2-C_2^2-C_3^2),
\\[1.5ex]\hspace*{-4em}
&=&\ds\frac{1}{4}(P_{13}+P_{24}-C_1-C_1^3),
\\[1.5ex]\hspace*{-4em}
Q_{13}^{(3,1)}=\left(Q_{31}^{(3,1)}\right)^{\dagger}
&=&\ds \frac{1}{8}\sum_{l=1}^4(-1)^l(D_l+D_l^2)+\frac{\I}{4}(C_2^2-C_3^2),
\\[1.5ex]\hspace*{-4em}
&=&\ds \frac{1}{4}(P_{13}-P_{24})-\frac{\I}{16}(P_{12}-P_{23}+P_{34}-P_{41})
\\[2ex]\hspace*{-4em}
&&\ds\mbox{}-\frac{\I}{16}(C_2+C_2^3-C_3-C_3^3),
\\[1.5ex]\hspace*{-4em}
\left.\begin{array}{r}
Q_{12}^{(3,1)}=\left(Q_{21}^{(3,1)}\right)^{\dagger}\\[1ex]
Q_{23}^{(3,1)}=\left(Q_{32}^{(3,1)}\right)^{\dagger}
\end{array}\right\}
&=&\ds\frac{1}{4}(1\pm\I)\sum_{l=1}^3\omega_4^{3l}D_l
+\frac{1}{4}(1\mp\I)\sum_{l=1}^3\omega_4^{l}D_l^2,
\\[1.5ex]\hspace*{-4em}
&=&\ds\frac{1}{4}(1+\I)(P_{12}-P_{34})+(1-\I)(P_{23}-P_{41})
\\[2ex]\hspace*{-4em}
&&\ds\mbox{}\mp\frac{1}{4}\Bigl[(1-\I)(C_2-C_2^3)+(1+\I)(C_3-C_3^3)\Bigr]. 
\end{eqnArray}
For the $\nu=(2,2)$ subspace, we have  
\begin{eqnArray}\nonumber
Q_{\lambda\lambda}^{(2,2)}&=&\ds
\frac{1}{12}(I+\sum_{k=1}^3C_k^2)\\[2ex]
&&\ds\mbox{}+\frac{\omega_3^{\lambda}}{12}(D_1+D_2^2+D_3+D_4^2)\\[2ex]
&&\ds\mbox{}+\frac{\omega_3^{2\lambda}}{12}(D_1^2+D_2+D_3^2+D_4),
\quad (\lambda=1,2)\\[1.5ex]
Q_{12}^{(2,2)}=\left(Q_{21}^{(2,2)}\right)^{\dagger}&=&
\ds\frac{1}{12}(C_3+C_3^3+P_{12}+P_{34})\\[2ex]
&&\ds\mbox{}+\frac{\omega_3}{12}(C_1+C_1^3+P_{13}+P_{24})\\[2ex]
&&\ds\mbox{}+\frac{\omega_3^{2}}{12}(C_2+C_2^3+P_{23}+P_{41}).
\end{eqnArray}
As we see there are many ways of representing the $Q_{k\ell}^{\nu}$s by linear
combinations of permutation operators, and here we choose the following
representation to see representations for the RFF basis operators in terms of
the Pauli spin operators.  

Define the hermitian operators $A^{\pm}_{j}$, $K_{j}$, and $L_{j}$ by 
\begin{eqnArray}\nonumber
\hspace*{-4em}
&A^{\pm}_{1}=P_{2134}\pm P_{1243},\quad 
A^{\pm}_{2}=P_{3214}\pm P_{1432},\quad 
A^{\pm}_{3}=P_{1324}\pm P_{4231},& \\[2ex] \nonumber
\hspace*{-4em}
&K_{1}=\I(P_{1342}-P_{1423}),\quad 
K_{2}=\I(P_{3241}-P_{4213}),&\\[1.5ex] 
\hspace*{-4em}
&K_{3}=\I(P_{2431}-P_{4132}),\quad 
K_{4}=\I(P_{2314}-P_{3124}),& \\[2ex] \nonumber
\hspace*{-4em}
&L_{1}=P_{2143},\quad L_{2}=P_{3412},\quad L_{3}=P_{4321},& 
\end{eqnArray}
where the $A^{\pm}_{j}$s are combinations of elements of the transposition
permutation subgroup,  
the $K_{j}$s are those of the cyclic permutation subgroup of order $3$, 
and the $L_{j}$ are those of the 2-cycle permutation operators. 
Their Pauli spin operator representations are
\begin{eqnArray}\nonumber
\hspace*{-4em}
&K_1=\ds-\frac{1}{2}\vec{\sigma}_{2}
\cdot (\vec{\sigma}_{3}\times\vec{\sigma}_{4}),
\quad K_2=\ds-\frac{1}{2}\vec{\sigma}_{3}
\cdot (\vec{\sigma}_{4}\times\vec{\sigma}_{1}),&\\[2ex]
&K_3=\ds-\frac{1}{2}\vec{\sigma}_{4}
\cdot (\vec{\sigma}_{1}\times\vec{\sigma}_{2}), \quad
K_4=\ds-\frac{1}{2}\vec{\sigma}_{1}
\cdot (\vec{\sigma}_{2}\times\vec{\sigma}_{3}),&\\[2ex]
&L_1=\ds\frac{1}{4}(I_{16}+\vec{\sigma}_1\cdot\vec{\sigma}_2)
(I_{16}+\vec{\sigma}_3\cdot\vec{\sigma}_4),&\\[2ex]
&L_2=\ds\frac{1}{4}(I_{16}+\vec{\sigma}_1\cdot\vec{\sigma}_3)
(I_{16}+\vec{\sigma}_2\cdot\vec{\sigma}_4),&\\[2ex]
&L_3=\ds\frac{1}{4}(I_{16}+\vec{\sigma}_2\cdot\vec{\sigma}_3)
(I_{16}+\vec{\sigma}_4\cdot\vec{\sigma}_1),&
\end{eqnArray}
besides the trivial ones for $A^{\pm}_{j}$. 
With these notations, we have 
\begin{eqnArray}
Q_{11}^{(4,0)}&=&\ds-\frac14 I_{16}+\frac16 (A_{1}^{+}+A_{2}^{+}+A_{3}^{+})
\\[2ex]&&\ds\mbox{}
+\frac{1}{12}(L_{1}+L_{2}+L_{3}),\\[3ex]
\left.\begin{array}{r}
Q_{11}^{(3,1)}\\[1ex]
Q_{33}^{(3,1)}
\end{array}\right\}
&=&\ds\frac{1}{4} I_{16}
\mp\frac{1}{8}(K_{1}+K_{2}+K_{3}+K_{4})-\frac{1}{4}L_{2},\\[4ex]
Q_{22}^{(3,1)}&=&
\ds\frac{1}{4}I_{16}-\frac{1}{4}(L_{1}-L_{2}+L_{3}),\\[4ex]
\hspace*{-2em}
\left.\begin{array}{r}
Q_{12}^{(3,1)}=\left(Q_{21}^{(3,1)}\right)^{\dagger}\\[1ex]
Q_{23}^{(3,1)}=\left(Q_{32}^{(3,1)}\right)^{\dagger}
\end{array}\right\}
&=&\ds\frac{1}{8}
\bigl[(1+\I)A_{1}^{-}+(1-\I)A_{3}^{-}\\[-1ex]&&\ds\mbox{}
\mp(\I K_{1}+K_{2}-\I K_{3}-K_{4})  
\bigr],\\[3ex]
Q_{13}^{(3,1)}=\left(Q_{31}^{(3,1)}\right)^{\dagger}&=&
\ds\frac{1}{4}(A_{2}^{-}+\I L_{1}-\I L_{3}),\\[3ex]
\nonumber
\left.\begin{array}{r}
Q_{11}^{(2,2)}\\[1ex]
Q_{22}^{(2,2)}
\end{array}\right\}
&=&\ds\frac{1}{4} I_{16}
-\frac{1}{12}(A_{1}^{+}+A_{2}^{+}+A_{3}^{+})\\&&\ds\mbox{}
\pm\frac{1}{8\sqrt{3}}(K_{1}-K_{2}+K_{3}-K_{4})\\[2ex] 
&&\ds\mbox{}+\frac{1}{12}(L_{1}+L_{2}+L_{3}),\\[3ex]
Q_{12}^{(2,2)}=\left(Q_{21}^{(2,2)}\right)^{\dagger}&=&
\ds\frac{1}{3}\bigl[A_{1}^{+}+\omega_{3}A_{2}^{+}
+\omega_{3}^{2}A_{3}^{+}\\[2ex] &&\ds\mbox{} 
-(L_{1}+\omega_{3}L_{2}+\omega_{3}^2 L_{3})\bigr]. 
\end{eqnArray}
This agrees with the result presented in section~IV~D of~\cite{JGB}. 
We remark that the RFF basis operators for the subspace $(2,2)$ $(j=0)$ 
can be expressed in terms of spin-0 singlet states. 
   
The MLOs are constructed from the diagonal elements with the proper choice of
coefficients.  
The most natural form among many different constructions is 
\begin{eqnArray}\hspace*{-6em}
1 \times Q_{11}^{(3,1)}+0 \times Q_{22}^{(3,1)}+(-1)\times Q_{33}^{(3,1)}&=&
\ds\frac{-1}{4} (K_{1}+K_{2}+K_{3}+K_{4})=M_{j=1},\\[2ex]\hspace*{-6em}
1 \times Q_{11}^{(2,2)}+(-1) \times Q_{22}^{(2,2)}&=&
\ds\frac{1}{\sqrt{12}}(K_{1}-K_{2}+K_{3}-K_{4})=M_{j=0}, 
\end{eqnArray}
and hence the CSCO, up to trivial multiplicative factors, are obtained as 
\begin{equation}
\mathrm{CSCO}_{\mathrm{sym}}=\bigl\{J^{2},J_{z},M_{j=1},M_{j=0}\bigr\}. 
\end{equation}
Note that this construction of the MLO for the ${j=1}$ subspace agrees with
\eref{eq:52} of section~\ref{sec:3-1}.

The corresponding angular momentum states in the $j=1$ subspace are 
\begin{eqnArray}
|1,1;\lambda\rangle&=&
\ds\frac{1}{2}\bigl(|1000\rangle\omega_4^{\lambda}
                    +|0100\rangle\omega_4^{2\lambda}
                    +|0010\rangle\omega_4^{3\lambda}
                    +|0001\rangle\bigr),\\[2ex] \nonumber
|1,0;\lambda\rangle&=&
\ds\frac{1}{\sqrt{8}}\bigl[
\bigl(|1001\rangle-|0110\rangle\bigr)(\omega_4^{\lambda}+1)\\
&&\ds\hphantom{\frac{1}{\sqrt{8}}\bigl[}
+\bigl(|0101\rangle-|1010\rangle\bigr)(\omega_4^{2\lambda}+1)
\\[1ex] &&\ds\hphantom{\frac{1}{\sqrt{8}}\bigl[}
+\bigl(|0011\rangle-|1100\rangle\bigr)(\omega_4^{3\lambda}+1)\bigr],\\[2ex] 
|1,-1;\lambda\rangle&=&\ds-\frac{1}{2}\bigl(
|0111\rangle\omega_4^{\lambda}
+|1011\rangle\omega_4^{2\lambda}
+|1101\rangle\omega_4^{3\lambda}
+|1110\rangle\bigr)
\end{eqnArray}
with ${\lambda=1,2,3}$, and those in the $j=0$ subspace are
\begin{eqnArray}
|0,0;\lambda\rangle
&=&\ds\frac{1}{\sqrt{6}}\bigr[
\bigl(|1001\rangle+|0110\rangle\bigr) \omega_3^{\lambda}
+\bigl(|0101\rangle +|1010\rangle\bigr)\omega_3^{2\lambda}\\[2ex]
&&\ds\hphantom{\frac{1}{\sqrt{6}}\bigr[}
+\bigl(|0011\rangle+|1100\rangle\bigr)\bigr] 
\end{eqnArray}
with ${\lambda=1,2}$.

From our construction, the symmetry of $\mathrm{CSCO}_{\mathrm{sym}}$ is clear. 
To this end, it is more convenient to introduce the following five conjugacy
classes of the permutation group $S_{4}$ based upon the disjoint cycles of the
subgroups by 
\begin{eqnArray}
Cl(1^{4})&=&\{ P_{1234} \},\\[1.5ex] 
Cl(21^{2})&=&\{ P_{2134}, P_{1243},P_{3214}, P_{1432}, P_{1324}, P_{4231} \},
\\[1.5ex] 
Cl(31)&=&\{ P_{1342},P_{1423}, P_{3241},P_{4213},P_{2431},P_{4132}, 
P_{2314},P_{3124} \},\\[1.5ex] 
Cl(4)&=&\{ P_{2341},P_{4123},P_{2413},P_{3142},P_{3421},P_{4312} \},\\[1.5ex] 
Cl(2^{2})&=&\{ P_{3412},P_{4321},P_{2143} \}. 
\end{eqnArray}
Then, beside a trivial symmetry under the identity $P_{1234}$, 
$M_{j=1}$ is invariant under the cyclic permutation subgroup of
order $4$ in (\ref{c4}), 
\begin{equation}
c_{4} M_{j=1} c_{4}^{-1}=M_{j=1}\quad\mbox{with\ } c_{4}\in C_{4},  
\end{equation}
and $M_{j=0}$ is symmetric under the $3$-cycle conjugacy class 
and the two $2$-cycle conjugacy classes, 
\begin{eqnArray}
c_{3}M_{j=0}c_{3}^{-1}&=&M_{j=0} \quad\mbox{with\ } c_{3}\in Cl(31),\\[1.5ex]
c_{2^{2}} M_{j=0}c_{2^{2}}^{-1}&=&M_{j=0} 
\quad\mbox{with\ } c_{2^{2}}\in Cl(2^{2}).  
\end{eqnArray}
We remark that $M_{j=1}$ is antisymmetric under $P_{3214}, P_{1423}\in Cl(21^2)$ 
and $P_{4321}, P_{2143}\in Cl(2^2)$, and $M_{j=0}$ is 
antisymmetric under the conjugacy classes $Cl(21^2)$ and $Cl(4)$. 
These symmetries should be contrasted with the standard binary coupling scheme
that gives $\mathrm{CSCO}_{12|34}$ or the like.
For identical spins, the MLOs of the kind $J^2_{12}$, $J^2_{34}$ are then only
invariant under two of the classes in $Cl(21^2)$ and one of the classes in
$Cl(2^2)$. 
This is our main conclusion:
The higher symmetry in the MLOs distinguishes this coupling scheme from all
other coupling schemes.

\section{Summary and discussion}\label{sec:SummDisc}
In this paper we have given a detailed analysis of the symmetric coupling of
four spin-1/2 systems by revealing the symmetry of the MLOs. 
The result is based on our proposal 
for the symmetric coupling of many identical angular momenta 
in which the Weyl-Schur duality plays a central role. 
The relation between the MLOs and the RFF subsystems is also clarified through
the analysis.  

An immediate extension is to provide the MLOs for the coupling of 
four arbitrary angular momenta without using the binary coupling. 
This long-standing open problem can be tackled by first solving 
the case of four identical angular momentum systems and then 
analyzing the eigenvalues of the obtained MLOs in the non-identical case. 

Another interesting problem is to study the possible coupling scheme 
for $N$ spin-1/2 systems in all subspaces along the ideas 
presented in this paper.

\ack
We would like to thank Gelo Tabia and Shiang Yong Looi for insightful
discussions. 
J~S is supported by MEXT. 
This work is supported by the National Research Foundation and the Ministry of
Education, Singapore.

\newcommand{\arXiv}[2][quant-ph]{(\textit{eprint}\ arXiv:#1/#2)}
\newcommand{\arxiv}[2][quant-ph]{(\textit{eprint}\ arXiv:#2 [#1])}

\end{document}